\documentclass[aps,prl,twocolumn,superscriptaddress,showpacs]{revtex4}

\usepackage{graphics}
\usepackage{epsfig}
\usepackage{color}
\begin{document}

\title{Fano resonances at light scattering by an obstacle
}
\author{Michael I. Tribelsky}

\affiliation{Moscow State Institute of Radioengineering, Electronics
and Automation (Technical University), 78 Vernadskiy Avenue, Moscow
119454, Russia}

\affiliation{Max-Planck-Insitut f\"ur Physik komplexer Systeme,
N\"othnitzer Str. 38, Dresden 01187, Germany}

\author{Sergej Flach}

\affiliation{Max-Planck-Insitut f\"ur Physik komplexer Systeme,
N\"othnitzer Str. 38, Dresden 01187, Germany}

\author{Andrey E. Miroshnichenko}

\affiliation{Nonlinear Physics Centre, Research School of
Physical Sciences and Engineering, Australian National University,
Canberra ACT 0200, Australia}
\affiliation{Max-Planck-Insitut f\"ur Physik komplexer Systeme,
N\"othnitzer Str. 38, Dresden 01187, Germany}

\author{ Andrey Gorbach}

\affiliation{Centre for Photonics and Photonic Materials,
Department of Physics, University of Bath, Claverton Down, Bath, BA2 7AY, UK}

\author{Yuri S. Kivshar}

\affiliation{Nonlinear Physics Centre, Research School of
Physical Sciences and Engineering, Australian National University,
Canberra ACT 0200, Australia}

\date{\today}

\begin{abstract}
It is shown that elastic resonance scattering of light by a finite-size 
obstacle with weak dissipation is analogous to quantum
scattering by a potential with quasi-discrete levels and exhibits
Fano resonances. Localized plasmons (polaritons), exited in the
obstacle by the incident light, are equivalent to the quasi-discrete
levels, while the radiative decay of these excitations plays exactly
the same role as tunnelling from the quasi-discrete levels for the
quantum problem. Mie scattering of light by a spherical particle and
an exactly solvable discrete model with nonlocal coupling
simulating wave scattering in systems with reduced spatial
dimensionality are discussed as examples.
\end{abstract}

\pacs{42.25.Bs,42.25.Fx,76.60.Gv,78.67.Bf}

\maketitle

{\em Introduction and motivation.} Light scattering by an obstacle is one of the
fundamental problems of electrodynamics, see, e.g.,
monographs~\cite{BH,BW}. Nowadays interest to this problem increases
even more than ever owing to its numerous applications in
subwavelength optics, information processing, nanotechnologies and
related fields~\cite{plasmons:Nature}. Recent studies of resonant
scattering by small particles with weak dissipation rates
(see~\cite{mitbsl06,mbvfnz05,zbwbslmhhyltcc04} and references
therein) have revealed new and unexpected features, namely the
inverse hierarchy of optical resonances, a complicated near-field
structure, unusual frequency and size dependencies etc., which allow
to name such a scattering \textit{anomalous}. However, despite the
highlighted peculiarities, the physical nature of both normal
resonant (Rayleigh) and anomalous scattering is the same and may be
described briefly as follows. Incident light excites localized
electromagnetic modes in a scattering particle (plasmons
or polaritons) oscillating with the frequency of the incident wave
$\omega$. The corresponding oscillations of polarization of the
particle result in an emission of electromagnetic waves with the
same frequency $\omega$ which constitute the scattered light.
Resonances in this picture correspond to the cases when $\omega$
occurs close to (the real part of) the eigenfrequency of one of the
localized modes.

\begin{figure}
 \includegraphics[width=90mm]{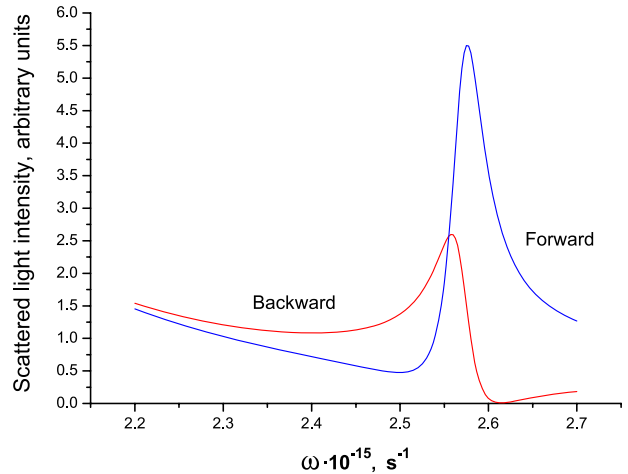}%
 \vspace*{-5mm}
\caption{(Color online) An example of typical Fano resonance profiles at elastic
light scattering. Forward (blue) and backward (red) scattering  by a
small spherical particle calculated according to the exact Mie
solution in the vicinity of the quadrupole resonance; the profiles
for $I_\parallel^{(s)}$ and $I_\perp^{(s)}$ are identical, cf. Eqs.
\protect{(\ref{I2par}), (\ref{I2per})}. The media and constants
model the sketch of experiment on the anomalous scattering proposed
in \protect{Ref.~\cite{JETP84}}, namely a colloidal potassium
particle with radius $R = 6.2\times 10^{-6}$ cm immersed in a
crystal of KCl. The refractive index of the crystal
$n_{\mbox{\tiny{KCl}}} = 1.5$; the dielectric permittivity of the
particle is approximated by the Drude formula: $\epsilon = 1 -
\omega_p^2/(\omega + i\nu)\omega$, where $\omega_p = 5.77\times
10^{15}$~s$^{-1}$ and $\nu = v_F/R$ is determined by collisions of
free electrons with the particle surface; $v_F = 10^8$~cm/s.}
\label{fig:fig2}
\end{figure}

In the present Letter we reveal that this physical
picture is analogous to the well-known case of \textit{Fano
resonances\/} in quantum physics~\cite{uf61}. This analogy sheds a
new light to the phenomenon. It allows to employ powerful methods
developed in the theory of the Fano resonances (such as, e.g., the
Feshbach-Fano partitioning theory) to describe the resonant light
scattering. It also explains easily certain features of the
anomalous scattering and related problems, namely sharp changes in
the scattering diagrams (from preferably forward to backward) upon
small changes in $\omega$, recently observed in numerical
simulations~\cite{APA} and a typical asymmetric shape of the resonance
lines (see Fig.~\ref{fig:fig2}) obtained in the present Letter by
analysis of the exact Mie solution of the light scattering problem
by a spherical particle~\cite{BW,BH}. We also introduce and study an
exactly solvable model accounting for all main features of the
resonance light scattering by a particle and exhibiting distinctive
Fano resonances. The model simulates wave scattering in an array of
coupled waveguides~\cite{Ablowitz}, photonic crystals~\cite{phcr}
and related systems. Remarkable similarity in manifestation of the
resonances in all these cases helps us revealing fundamental links
between these phenomena.

It should be stressed that though the Fano resonances in light
scattering have been discussed before, attention has been paid
generally to microscopic aspects of inelastic scattering in a simple
plane geometry (see, e.g., Ref.~\cite{Falk}). In contrast to that, we
focus on elastic light scattering by a finite-size particle when
diffraction plays a fundamental role in the scattering process. The
scattering is accompanied by a two-step transformation: incident
plane wave $\rightarrow$ localized electromagnetic modes
$\rightarrow$ scattered light. On account of the second step of this
transformation (radiative damping) the localized modes have a finite
life-time, being actually \textit{quasilocalized\/} even at zero
dissipation rate. Note however, that as long as the scattering of a
continuous wave is concerned, the losses of energy by the
quasilocalized modes are exactly compensated by the gain from the
incident wave. Then, the amplitude of the modes becomes
time-independent.

Another route for the incident light is just to bypass the
scatterer, as well as to excite (nonresonantly) localized modes
whose eigenfrequencies lie beyond the close vicinity of $\omega$.
Interference of the incident and re-emitted light generates a
complicated near-field structure and may give rise either to strong
enhancement (constructive interference) or strong suppression
(destructive interference) of the electromagnetic field. The analogy
to Fano resonances of a quantum particle scattered by a potential
with quasidiscrete levels becomes straightforward. The two possible
routes for the incident light correspond to the resonant and direct
(background) scattering, while the radiative decay of the
(quasi)localized modes is identical to tunnelling from the
quasidiscrete level.

{\it Light scattering by a small spherical particle}. Though the
analogy mentioned is pretty general and valid for any particle size
and shape, for the sake of simplicity in what follows we consider
light scattering by a small spherical particle described by the
exact Mie solution~\cite{BW,BH}. According to the solution for a
plane polarized wave propagating along $z$-axis with vector {\bf E}
parallel to $x$-axis the intensities of waves scattered in a given
direction are described by the expressions
\begin{equation}\label{I}
    S_\parallel^{(s)} = I_\parallel^{(s)}\cos^2 \varphi; \;\;
  S_\perp^{(s)} = I_\perp^{(s)}\cos^2 \varphi,
\end{equation}
where the subscripts indicate the corresponding polarization
(relative to the incident wave), and up to a certain common
multiplier 
\begin{eqnarray}
  I_\parallel^{(s)}\!\! & \propto & \!\! \left| \sum_{\ell=1}^\infty
  \frac{2\ell+1}{\ell(\ell+1)}\left(a_\ell P_\ell^{(1)'}\!(\cos
  \theta)\sin \theta - \right.\right.\nonumber\\
  & &\hspace*{25mm} \left.\left.- b_\ell\frac{P_\ell^{(1)}\!(\cos \theta)}{\sin
  \theta}\right)\right|^2,\label{Ipar}\\
  I_\perp^{(s)}\!\! & \propto & \!\! \left| \sum_{\ell=1}^\infty
  \frac{2\ell+1}{\ell(\ell+1)}\left(a_\ell\frac{P_\ell^{(1)}\!(\cos \theta)}{\sin
  \theta} - \right.\right.\nonumber\\
  & &\hspace*{25mm} \left.\left.- b_\ell P_\ell^{(1)^{\prime}}\!(\cos
  \theta)\sin \theta\right)\right|^2.\label{Iper}
\end{eqnarray}
Here $\theta$ and $\varphi$ are the polar and azimuthal angles of
the spherical coordinate frame whose center coincides with that of
the particle and $z$-axis with the propagation direction of the
incident wave; $P_\ell^{(1)}\!(\cos \theta)$ stands for the
associated Legendre polynomial, $P_l^{(1)^{\prime}}\!(\cos \theta) = d
P_l^{(1)}\!(\cos \theta)/d(\cos \theta)$, scattering amplitudes
$a_l$ and $b_l$ may be written in the following form:

\begin{equation}\label{ab}
   a_\ell = \frac{F^{(a)}_\ell  \left( {q ,\epsilon}
\right)}{F^{(a)}_\ell  \left( {q ,\epsilon} \right) + iG^{(a)}_\ell
\left( {q ,\epsilon} \right)},
\end{equation}
and similar for $b_\ell$ with replacement $F^{(a)}, G^{(a)}
\rightarrow F^{(b)}, G^{(b)}$; where $\epsilon(\omega)$ is the
relative (with respect to the environment) dielectric permittivity
of the particle, $q = kR$, $k$ stands for the wavenumber of the
incident wave in vacuum and $R$ for the particle radius. As for
$F^{(a,b)},\;\; G^{(a,b)}$, they are expressed in terms of the
Bessel [$J_{l+1/2}(\zeta)$] and Neumann [$N_{l+1/2}(\zeta)$]
functions. The corresponding expressions are cumbersome and may be
found in Ref.~\cite{zbwbslmhhyltcc04}. The partial scattering
cross-section ($\sigma_{sca}^{(\ell)}$), is connected with
$a_\ell,\; b_\ell$:
\begin{equation}\label{sigma}
\sigma_{sca}^{(\ell)} = \frac{2 \pi}{k^2}(2\ell + 1)(|a_\ell|^2 +
|b_\ell|^2).
\end{equation}

We remind briefly results of the analysis of the Mie solution (for
more details see, e.g., \cite{mitbsl06} and references therein).
Optical resonances are defined by the condition $G^{(a,b)}_\ell
\left( {q ,\epsilon_\ell} \right)=0$. For a small particle ($q \ll
1$) the condition $G^{(b)}_\ell \left( {q ,\epsilon_\ell}
\right)=0$, regarded as an equation for $\epsilon_\ell$, does not
have any real solutions, while solutions of the equation $G^{(a)}_l
\left( {q ,\epsilon_\ell} \right)=0$ have the form $\epsilon_\ell =
-(1+\ell)/\ell + O(q^2).$ Through the dependence
$\epsilon(\omega)$, these solutions define resonant frequencies
$\omega_\ell$ for non-dissipating materials. For weakly dissipating
materials the roots of the equation $G^{(a)}_\ell \left( {q
,\omega_\ell} \right)=0$ have small imaginary parts. In this case
the resonant frequencies are equal to the real parts of the
corresponding roots.  Next, at $q \ll 1$ amplitudes $b_\ell$ are
small relative to $a_\ell$ and may be dropped. Regarding $F^{(a)}$,
in this limit $F^{(a)}_\ell = O\left(q^{2\ell+1}\right)$ and it does
not vanish anywhere but at the trivial point $\epsilon =1$; as for
$G^{(a)}$, away from the close vicinity of the resonance frequencies
it is of order of one. All together that result in sharp optical
resonances for $\sigma_{sca}^{(\ell)}$, with a typical symmetric
Lorenzian (Breit-Wigner) profile~\cite{n2}.

However, such a profile is not always the case for
$I^{(s)}_{\parallel,\perp}$. Let us first discuss the
non-dissipative limit (Im$\,\epsilon =0$). Considering excitations
of non-resonant modes as background scattering and bearing in mind
the narrowness of the resonance lines in this case~\cite{mitbsl06}
we obtain that the corresponding amplitudes $a_\ell \approx a_{l0}$,
where $a_{l0} = O\left(q^{2l+1}\right)$ is a purely imaginary
quantity equal to $ - i
F^{(a)}_\ell(q,\epsilon)/G^{(a)}_\ell(q,\epsilon)$. Then, the dipole
mode ($\ell=1$) always plays the dominant role in the background
scattering. Accordingly, in the vicinity of the dipole resonance the
intensity of the dipole scattering always remains large relative to
the contribution of the other modes. As a result interference of the
resonant scattering with the background gives rise to only small
corrections to the resonant mode and the profiles
$I^{(s)}_{\parallel,\perp}(\epsilon)$ remain Lorenzian.

For higher-order resonances with $\ell>1$, which are also well
pronounced in the non-dissipative limit even for a small
particle~\cite{mitbsl06}, the situation is qualitatively different.
In this cases the resonance scattering arises from the corresponding
background, whose amplitude $a_{\ell0}$ is {\it much smaller\/} than
that for the background dipole mode ($a_{10}$). As $\epsilon$
approaches $\epsilon_\ell$ the resonant amplitude rapidly increases
and becomes much greater than $a_{10}$, reaching a maximum at
$\epsilon = \epsilon_\ell$ and decreasing back to $a_{\ell0}$ upon
further change of $\epsilon$. Thus, in the vicinity of
$\epsilon_\ell$ there are two points (one to the left, the other to
the right from $\epsilon_\ell$) where $|a_{\ell}|=|a_{10}|$.
Therefore, in addition to the constructive interference of the
resonant mode with the background dipole scattering at $\epsilon =
\epsilon_\ell$, we also find annihilation of the two modes at one of
the equal-amplitude-points (passing the resonance adds $\pi$ to the
phase of the resonant mode, so only one of the two points satisfies
the necessary phase condition for the destructive interference with
the background mode). In this case an asymmetric Fano resonance
profile is obtained.

To illustrate this reasoning let us consider the vicinity of the
quadrupole resonance ($\ell = 2$).
Taking $F^{(a)}_{1,2},\; G^{(a)}_1$ at
$\omega=\omega_\ell$, employing expansions of the functions in
powers of $q$~\cite{mitbsl06} and additionally expanding $G^{(a)}_2$
in powers of $\delta\epsilon = \epsilon~-~\epsilon_2$ [we remind
that $G^{(a)}_2(q,\epsilon_2)=0$] one obtains the following approximate
expressions for $I^{(s)}$:
\begin{eqnarray}
  I^{(s)}_\parallel &\propto& \left| i q^3 \cos\theta +
    \frac{q^5\cos2\theta}{2(q^5 -12i\delta\epsilon)}\right|^2,\label{I2par} \\
  I^{(s)}_\perp &\propto& \left| i q^3 +
    \frac{q^5\cos\theta}{2(q^5 -12i\delta\epsilon)}\right|^2.\label{I2per}
\end{eqnarray}
There are two characteristic scales in Eqs.~(\ref{Ipar}), (\ref{Iper}),
describing (i) a sharp resonant lineshape centered at $\delta\epsilon =
0$ with width $\Gamma = q^5/6$~\cite{n4} and (ii) full suppression
of the scattering which occurs at much greater scale:
$\delta\epsilon_\parallel = -q^2\cos2\theta/24\cos\theta$ and
$\delta\epsilon_\perp = -q^2\cos\theta/24$, but still inside the
resonance region ($|\delta\epsilon_{\parallel,\perp}| \ll 1$). Note
also that there is change of sign of
$\delta\epsilon_{\parallel,\perp}$ with variation of $\theta$
shifting of the annihilation point from one side of the resonance
peak to another.

Finite dissipation and/or increase in $q$ will broaden the
resonances. It results in an overlap of resonance lines of different
orders, which eventually may produce very complex profiles.
Nevertheless the phenomenon remains qualitatively the same as long
as the dissipation is weak. Profiles of forward and backward
scattering, calculated for a potassium colloidal nanoparticle
immersed in a KCl crystal, are presented in Fig.~\ref{fig:fig2} as
an example. The calculations are performed for a realistic
dependence $\epsilon(\omega)$ fitting actual experimental data.
Note, that in accordance with Eqs.~(\ref{I2par}), (\ref{I2per}) the
points of destructive interference for the forward ($\theta = 0$)
and backward ($\theta = \pi$) scattering lie on different sides of
the corresponding resonant peaks~\cite{n3}. 

Now let us inspect one-dimensional resonant wave scattering
in systems with reduced spatial dimensionality. Numerous examples of
such systems could be found, e.g., in book~\cite{book}. Fano
resonances exhibited by these systems have exactly the same nature
as those discussed above. To make it sure we consider a localized
point defect in a simple 1D discrete chain with interaction between
the nearest and next-to-nearest neighbor sites (the nonlocal
coupling is a fundamental feature of the model).
%
\begin{figure}
 \includegraphics[width=60mm]{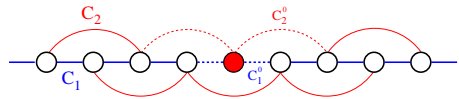}%
 \caption{\label{fig:fig3}
(Color online) Schematic view of a discrete 1D chain of sites with a defect
described by Eq.~(\ref{eq:eq1}). Solid and dashed lines correspond
to the resonant and background scattering, respectively, see the
text for more details.}
\end{figure}
%
The model is described by the following set of equations (see also
Fig.~\ref{fig:fig3}):
\begin{equation}\label{eq:eq1}
   i\frac{d\phi_n}{dt}\! +\!\!\! \sum_{j=-2}^2\!\!\!C_{n+j}^{n}\phi_{n+j}\!=\!0;\; C_n^n
   \!=\!
   \omega_0\delta_{n0}\phi_n;\; C_n^m = C_m^n,
\end{equation}
where the integer $n$ labels the spatial sites. All sites of the
chain but a defect situated at $n=0$ are identical. Accordingly,
$C_{n\pm1}^n = 1,\; C_{n\pm2}^n = \gamma$ for $|n|>2$. As for the
defect, $C_{\pm1}^0 = \mu,\; C_{\pm 2}^0 = \mu\gamma$ and $\omega_0$
stands for the shift of the defect intrinsic frequency with respect
to that for the other sites of the chain.



The solvability condition for a linear travelling wave solution
[$\phi \sim \exp(i\omega_q t - iqn)$] in the defect-free region of
Eq.~(\ref{eq:eq1}) results in the following dispersion relation:
\begin{eqnarray}\label{eq:eq3}
\omega_q = 2(\cos q+\gamma\cos 2q),
\end{eqnarray}
where $q$ is a (quasi)wave vector. To keep the one-to-one
correspondence between $\omega_q$ and $q$ from the non-trivial range
$0 \leq q \leq \pi$ we restrict $|\gamma| \leq 1/4$.

The defect-free region allows for another type of solutions whose
permitted frequencies cover the entire band of those for
Eq.~(\ref{eq:eq3}): $-2(1 - \gamma) \leq \omega_q  \leq 2(1 +
\gamma)$. These solutions are obtained by assigning complex values
to $q$ in Eq.~(\ref{eq:eq3}) $q = \pi \pm i p;\; p \geq 0$ and
correspond to exponentially decaying in space localized states.

To study wave scattering by the defect in our model we consider
an incident wave with amplitude equals one advancing from the left
~\cite{n1}. Then, the corresponding boundary conditions read:
\begin{eqnarray}\label{eq:eq4}
\phi_n=e^{i\omega_{q}t}\left\lbrace
              \begin{array}{lc}
                 e^{-iqn}+\rho_1e^{iqn}+\rho_2e^{pn}\;,&n<-2\\
                 \tau_1e^{-iqn}+\tau_2e^{-pn}\;,&n>2\;,
               \end{array}
        \right.
\end{eqnarray}
where $\tau_1$ and $\rho_1$, $\tau_2$ and $\rho_2$ are the
transmission and reflection amplitudes of the propagating and
evanescent waves from both sides of the defect, respectively.
The transmission coefficient $T$ is
related to $\tau_1$ by the expression $T = |\tau_1|^2$.

The linear problem Eqs.~(\ref{eq:eq1})-(\ref{eq:eq4}) is exactly
solvable, but the solution is extremely cumbersome. It is
remarkable, however, that the expression for $\tau_1$
has a structure identical to that of Eq.~(\ref{ab}),
namely
\begin{equation}\label{tau}
    \tau_1 = -\frac{F^{(\tau_1)}}{F^{(\tau_1)} + i G^{(\tau_1)}},
\end{equation}
where $F^{(\tau_1)}$ and $G^{(\tau_1)}$ are real quantities.
Eq.~(\ref{tau}) is simplified drastically at small $\gamma,\; \mu$
and $\omega_0$ under the additional restriction of weak coupling
between the defect and the chain $\mu^2 \ll |\omega_0\gamma|$. In
this case
\begin{equation}\label{tau1app}
    \tau_1 \approx -\frac{\mu^2 + \gamma\delta\omega}{\mu^2 +
    \gamma\delta\omega -i\delta\omega/2},
\end{equation}
where $\delta\omega = \omega_q - \omega_0$. The corresponding
$T(\delta\omega)=1$ at $\delta\omega=0$, i.e., when $\omega_q$
coincides with the intrinsic defect frequency and vanishes at
$\delta\omega = -\mu^2/\gamma$, cf. the discussed above resonance
scattering of light.

To make sure that the obtained properties of the transmission
coefficient are related to the resonant excitation of the defect,
we calculate $\phi_0$. In the same approximation (small
$\gamma,\; \mu,\;\omega_0$ and $\mu^2 \ll |\omega_0\gamma|$) we find:
\begin{equation}\label{phi0}
    \phi_0 \approx -\frac{\mu}{\mu^2 +
    \gamma\delta\omega -i\delta\omega/2}.
\end{equation}
Thus, $|\phi_0|^2$ exhibits a typical sharp Lorenzian profile
centered about $\delta\omega \approx -4\gamma\mu^2$.

More extensive analysis of Eqs. (\ref{eq:eq1})-(\ref{eq:eq4}) free
from the imposed restrictions for $\gamma,\; \mu$ and $\omega_0$
does not add any qualitative difference to the results presented
here, except for the fact, that the locations of resonant
transmission and reflection may be separated strongly in the frequency
space.
Note, though the next-to-nearest sites interaction is crucial to the
model, its generalization to more extended interactions (provided
the coupling constants become smaller with increase in distance
between the interacting sites) gives rise just to small quantitative
corrections to the model characteristics.

{\it Conclusions}. A deep connection between the light scattering by
a finite obstacle and Fano resonances in quantum physics is revealed
and illustrated by the analysis of the exact Mie solutions and a one-dimenssion
model. The concept developed in the present Letter may be
generalized to more complicated problems, e.g., the scattering by a
system of particles including coherent backscattering and weak
localization phenomena, scattering in waveguides, etc. In each
particular system, the specific resonant states are different but
all of them act in the same way being described by the similar physics.

{\it Acknowledgements}. We are grateful to B. Rubinstein for a help
with applications of {\it Mathematica\/} to the analysis of Eqs.
(\ref{eq:eq1})-(\ref{eq:eq4}) and to K. Kikoin for
discussions of our results. This work has been partly supported by
the Australian Research Council.


\begin{thebibliography}{11}
\bibitem{n2}We should stress this is true only for small particles. If a
particle is not small the line $\sigma_{sca}^{(\ell)}(\omega)$ may
also have a typical asymmetric Fano resonance shape.
\bibitem{n4}The obtained expression for $\Gamma$ coincides with that
given by Eq. (8) in Ref.~\cite{mitbsl06} for the line width of the
anomalous scattering. Note that there is a
misprint in Eq. (8), Ref.~\cite{mitbsl06} (missed factor 2 in the
numerator).
\bibitem{n3} A minor shift of the resonance peak with variation in
$\theta$ is caused by the mentioned effects of finite $q$ and finite
dissipation.
\bibitem{n1}The direction of the wave propagation is
given by sign of the group velocity $d\omega_q/dq$ which is positive
at $q > 0$ and $\gamma \leq 1/4$, see Eq.~(\ref{eq:eq3}).
\end{thebibliography}
\end{document}